# Sub-Micron Gate β-$Ga_2O_3$ FinFETs with Current Density > 750 mA/mm and $I_{ON}/I_{OFF}$ Ratio > $10^9$

Nabasindhu Das[1], Cameron Gorsak[2], Pushpanshu Tripathi[2], Ankita Kashyap[1], Noor Jahan Nipu[3], Advait Gilankar[1], Hari Nair[2], Uttam Singisetti[3], Trevor Thornton[1], Nidhin Kurian Kalarickal[1]

[1] *School of Electrical, Computer and Energy Engineering, Arizona State University, Arizona, USA*

[2] *Department of Materials Science & Engineering, Cornell University, Ithaca, New York, USA*

*3Department of Electrical Engineering, University at Buffalo, New York, USA*

*ndas11@asu.edu, Phone: (602)-813-6289*

In this work, we demonstrate submicron-gate β-$Ga_2O_3$ FinFETs that integrate Si δ-doped channels with selectively regrown $n^{++}$ source/drain contacts. The devices employ 100 nm-wide fins to improve electrostatic control of a high sheet charge density channel of ~$3.3x10^{13}$ $cm^{-2}$. A 150 nm-gate-length FinFET exhibits a maximum drain current density of 764 mA/mm at $V_{GS}$ = 1 V and $V_{DS}$ = 10 V, an on-resistance of 9.1 Ω·mm, and an $I_{ON}/I_{OFF}$ ratio exceeding $10^9$. The device also shows a peak transconductance of ~43 mS/mm and de-embedded $f_T/f_{MAX}$ values of 6.3/2.0 GHz. Pulsed I-V measurements indicate ~20% current collapse, attributed primarily to trapping at etched fin sidewalls and the $Al_2O_3$/ β-$Ga_2O_3$ interface. These results show that δ-doped β-$Ga_2O_3$ FinFETs can simultaneously achieve high drive current, strong current modulation, and GHz-frequency operation, providing a scalable device platform for future power and RF electronics.

β-$Ga_2O_3$ has emerged as a compelling ultra-wide-bandgap semiconductor for next-generation power and RF electronics owing to its large bandgap (~4.8 eV), high critical electric field (~8 MV/cm), reasonable electron transport properties (μ=200 $cm^2$/V-s, $v_{sat}$=1.5-2 x $10^7$ cm/s), strong n-type doping capability, and the availability of native bulk substrates [1], [2], [3]. For high frequency transistors, vertically scaled channels with high mobility and large sheet charge density are essential to achieve high current density and enhanced $f_T/f_{MAX}$[4]. However, uniformly doped channels commonly employed in β-$Ga_2O_3$ FETs are not ideal for aggressive scaling due to incomplete channel depletion and pronounced short-channel effects. Within this material system, Si δ-doping has emerged as an effective channel-engineering approach, enabling a vertically scaled channel with controllable high sheet charge while preserving comparatively strong carrier transport by confining the ionized donors to an ultrathin sheet [5], [6]. Earlier demonstrations of β-$Ga_2O_3$ δ-doped films and FETs established narrow doping profiles, low-resistance regrown source/drain contacts, and GHz-class RF operation, confirming the scalability of this approach [7], [8]. However, when such high sheet charge density channels are implemented in planar device geometries, further scaling becomes increasingly limited by the difficulty of fully depleting the channel due to short-channel electrostatics. FinFET architecture offers a viable pathway to address these challenges by enhancing electrostatic control through multi-sided gate coupling, therefore suppressing short-channel effects in high charge density channels. Furthermore, the addition of the two gate sidewalls enables lateral depletion of the channel, resulting in lower electric fields at the gate junction when the channel is fully depleted. Early β-$Ga_2O_3$ wrap-gate FinFETs demonstrated that lateral depletion of fins could enable enhancement-mode operation, although low on-current and substantial sidewall/interface trapping limited their performance [9]. Bae et al. subsequently reported 50-nm-wide exfoliated nanomembrane tri-gate FETs with a performance of 350 mA/mm current, a 61 mV/dec subthreshold swing, and an $I_{ON}/I_{OFF}$ ratio of $1.5 \times 10^9$ [9]. An epitaxially grown channel with tri-gate MESFETs later achieved 187 mA/mm on current and a power figure of merit of 0.95 GW/cm² [10], while 0.35-μm-gate tri-gate FinFETs reached $f_T/f_{MAX}$ values of 5.4/11.4 GHz [11]. In parallel, studies were done to suppress etch-induced sidewall damage resulting in plasma-free metal-assisted chemical etching produced high-quality fin sidewalls [10] and FinFETs with only 9.7 mV transfer hysteresis [12]. More recently, our previous work demonstrated 100-nm Si δ -doped fins modulating a sheet charge of $3.3 \times 10^{13} cm^{-2}$ and delivering 410 mA/mm on-state current density with $f_T/f_{MAX}$ values of 3.8/2.1 GHz [11]. Collectively, these reports demonstrate the electrostatic and RF advantages of multi-gate β-$Ga_2O_3$ channels while identifying fin-sidewall and gate-stack quality as central performance constraints [13], [14], [15]. In this work, we demonstrate submicron β-$Ga_2O_3$ FinFETs (Fig. 1 (a), (c)) that integrate an all MOCVD process including high-sheet-charge Si δ-doped channel and selectively regrown contacts. This approach enables strong modulation of >3 $\times 10^{13}$ $cm^{-2}$ channel charge, achieving drain current densities exceeding 750 mA/mm and $I_{ON}/I_{OFF}$ current ratios greater than $10^9$.

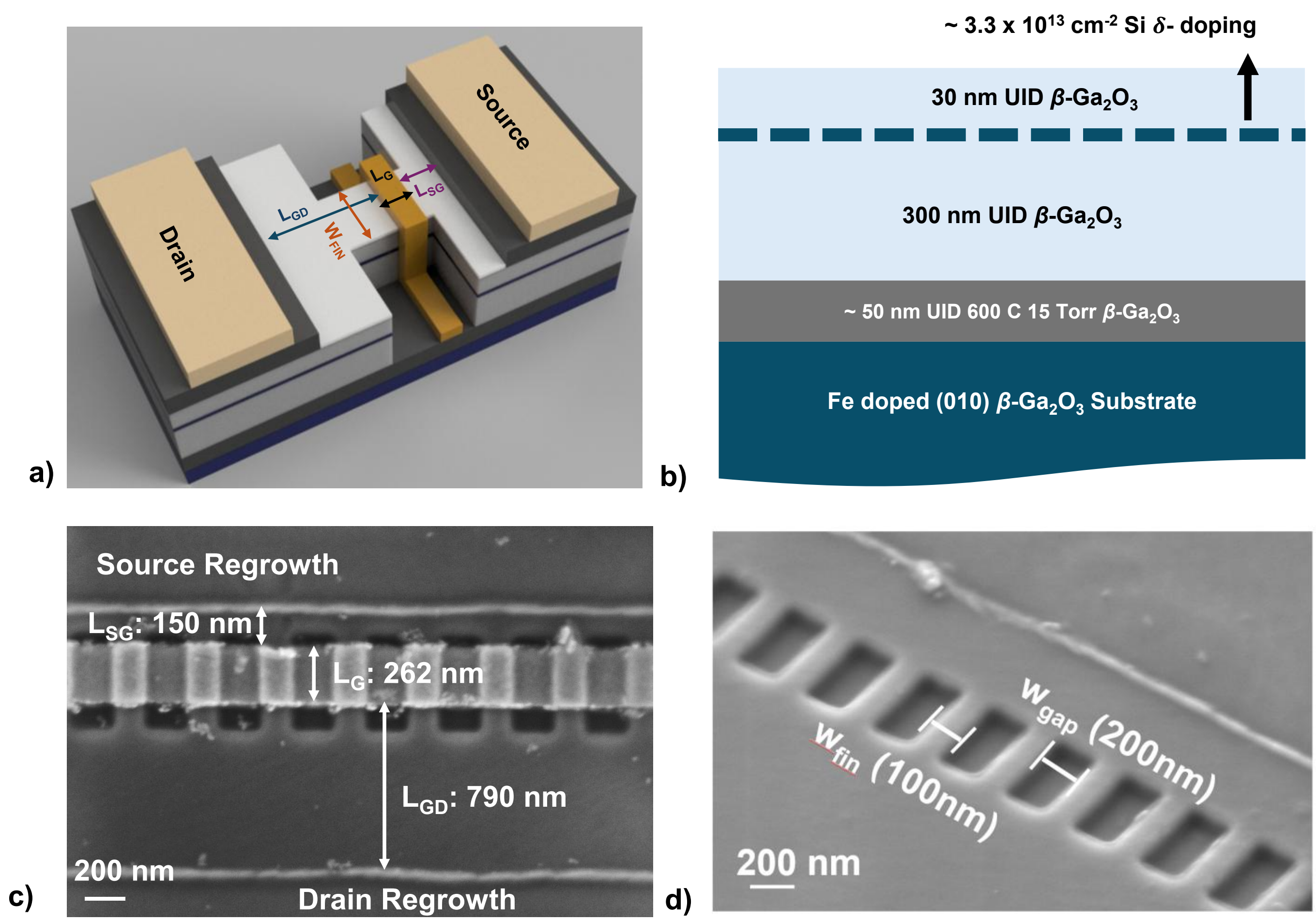

Fig 1. a) Schematic of β-$Ga_2O_3$ submicron delta-doped FinFET b) MOCVD grown epitaxial layer structure with delta-doped channel c) Tilted SEM image showing conformal metal deposition d) Tilted SEM Image showing fins with 100 nm width and 200 nm spacing.

The δ-doped β-$Ga_2O_3$ epitaxial structure used in this work is illustrated in Fig. 1(b). The layers were grown by metal–organic chemical vapor deposition (MOCVD) in an Agnitron Agilis 100 reactor on an Fe-doped (010) β-$Ga_2O_3$ substrate. To suppress the influence of the Fe surface riding effect [16], an initial ~50 nm unintentionally doped (UID) β-$Ga_2O_3$ layer was grown at a low temperature of 600 °C and 15 Torr using Ar as the carrier gas, with triethylgallium (TEGa) and $O_2$ as the gallium and oxygen precursors, respectively. This was followed by a ~300 nm UID β-$Ga_2O_3$ buffer layer. SIMS characterization of samples grown during the same timeframe revealed an unintentionally high Si background of ~$1\text{x}10^{17}$ $cm^{-3}$ in the buffer layer. The δ-doped sheet was subsequently formed through pulsed silane exposure (~20 nmol $SiH_4$) following a TEGa purge to ensure abrupt dopant incorporation. Finally, a ~30 nm UID β-$Ga_2O_3$ cap layer was grown to complete the structure. The resulting Si δ-doped sheet density was approximately $3.3\text{x}10^{13}$ $cm^{-2}$. The fin fabrication process was carried out using electron-beam lithography (EBL) followed by low-power Inductively Coupled Plasma Reactive Ion Etching (ICP-RIE) etching. A combination of dielectric and electron beam resist bilayer mask strategy was employed to achieve 100 nm feature definition while maintaining sufficient etch selectivity for anisotropic pattern transfer. Since high-resolution EBL requires a thin resist layer, a 10 nm $SiO_2$ hard mask was first deposited by plasma-enhanced chemical vapor deposition (PECVD) to support the subsequent ~90 nm deep β-$Ga_2O_3$ etch. An 80 nm layer of ZEP520A resist was then spin-coated and patterned using a 100 keV EBL system (Elionix ELS-BODEN). Pattern transfer into the β-$Ga_2O_3$ was performed using a low-power $BCl_3$-based ICP-RIE etch. The etch conditions ($BCl_3$ flow: 20 sccm, ICP power: 20 W, RF bias power: 15 W, pressure: 5 mTorr) were carefully optimized to minimize ion bombardment and plasma-induced damage to the fin sidewalls and the δ-doped channel. Suppression of surface damage is critical in submicron β-$Ga_2O_3$ fins, where sidewall states and defect generation can severely degrade carrier mobility. The low bias etch chemistry enabled reproducible, anisotropic 100 nm wide fins with 200 nm gap (Fig. 1(d)). No post-etch surface treatment was performed to further mitigate etch plasma damage.

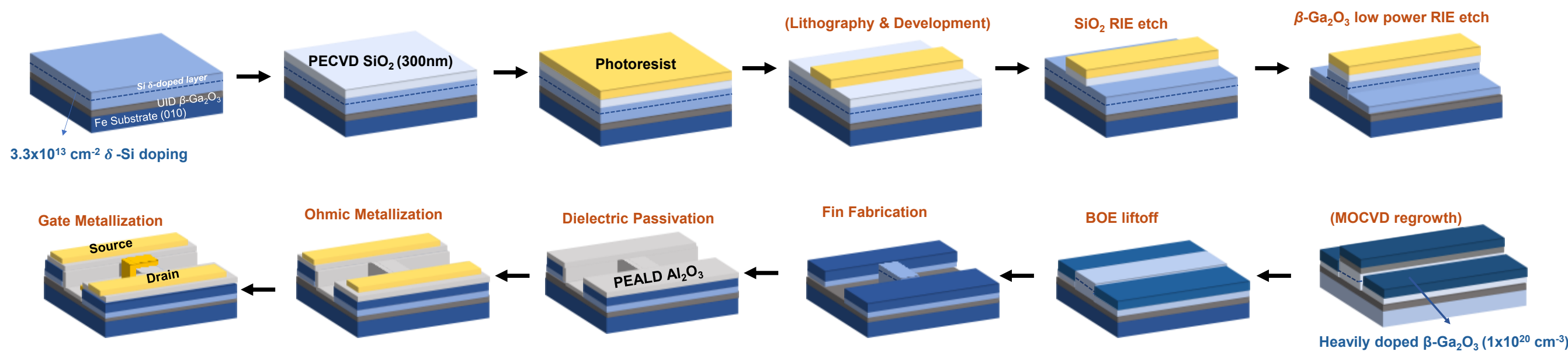

Fig.2 Process flow highlighting key fabrication steps integrating MOCVD-grown Si δ-doped channels, low-damage $BCl_3$-based fin etching, selective $n^+$ source/drain regrowth, ALD $Al_2O_3$ gate dielectric formation, and conformal Ni gate metallization.

Low-resistance source/drain contacts were formed using a selective $n^{++}$ ohmic regrowth process[7]. The source and drain regions were first patterned using a thick PECVD deposited $SiO_2$ hard mask to define regrowth windows. The exposed β-$Ga_2O_3$ surface was then etched to a depth of approximately 60 nm using a low-power $BCl_3$-based inductively coupled plasma reactive ion etching (ICP-RIE) process. The low bias etch conditions were used to minimize plasma-induced surface damage prior to regrowth and preserving the interface quality. Subsequently, heavily doped $n^{++}$ β-$Ga_2O_3$ ($N_D$-$N_A$ ≈ 1x$10^{20}$ $cm^{-3}$) was selectively regrown by MOCVD to form the source/drain access regions. Excess β-$Ga_2O_3$ deposited outside the regrowth windows was removed by BOE liftoff of the $SiO_2$ hard mask. This selective $n^{++}$ regrowth strategy, combined with low-damage etching prior to regrowth, is critical for minimizing interfacial resistivity and directly enables the low contact resistance observed in the fabricated devices. A mesa etch of approximately 200 nm was then performed to provide electrical isolation between adjacent devices. Following isolation, an 8 nm $Al_2O_3$ gate dielectric was deposited by plasma-enhanced atomic layer deposition (PEALD). The use of PEALD enables improved dielectric film density and interface quality, which are essential for suppressing gate leakage and maintaining strong electrostatic control in submicron FinFET geometries[17]. The $Al_2O_3$ dielectric was selectively removed from the source/drain contact regions prior to metallization. Ohmic contacts were formed by deposition of a Ti/Au (30/30 nm) metal stack using e-beam evaporation, followed by rapid thermal annealing at 470 °C for 1 min in $N_2$ ambient to promote alloying. Finally, an 80 nm Ni Schottky gate was defined by e-beam lithography and deposited by e-beam evaporation with 180° sample rotation to ensure uniform metal coverage along the fin sidewalls. The complete process flow is shown in Fig 2.

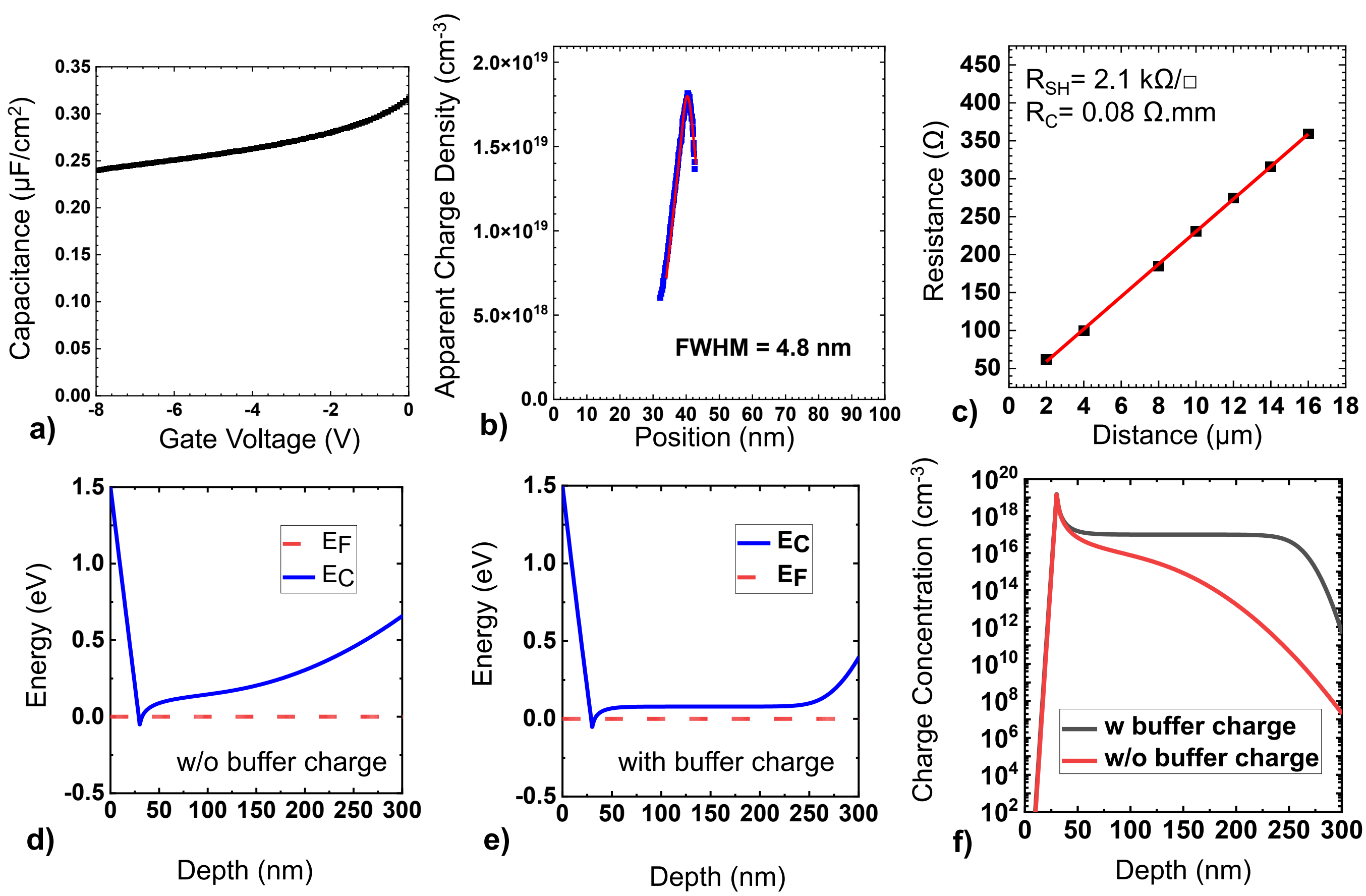

Fig. 3 a) Capacitance-Voltage profile for the delta-doped planar MOSCAP b) Apparent carrier density profile extracted from CV measurements c) TLM measurements showing sheet resistance and total contact resistance d) -f) Simulated energy band diagrams *(left, center)* and corresponding carrier concentration profiles *(right)* as a function of depth for the β-$Ga_2O_3$ structure without and with an unintentionally doped buffer layer, illustrating the impact of buffer charge on band bending, Fermi level position, and the resulting channel charge distribution.

Hall-effect measurements conducted on isolated van der Pauw structures yield a δ-doped sheet carrier density of ~$3.3 \times 10^{13}$ $cm^{-2}$ and a room-temperature electron mobility of 89 $cm^2$ $V^{-1}s^{-1}$. Capacitance–voltage (C–V) measurements (Fig. 3(a)) were performed on planar MOS structures using an Agilent B1500 parameter analyzer at a frequency of 1 MHz. The extracted apparent carrier density profile (Fig. 3(b)) exhibits a narrowly confined scaled doping profile with a peak at 40 nm and a full width at half maximum (FWHM) of 4.8 nm[18]. Due to the large sheet charge density, the planar MOS structures showed high leakage beyond -10 V, therefore C-V measurements were limited to -10 V. Transfer length method (TLM) measurements (Fig. 3(c)) were performed on structures with a width of 100 μm, yielding a contact resistance of 0.08 Ω·mm and a sheet resistance of 2.1 kΩ/□. The sheet resistance extracted from TLM measurements shows agreement with that independently estimated from Hall-effect measurements (2.1 kΩ/□), demonstrating consistency between the two techniques.

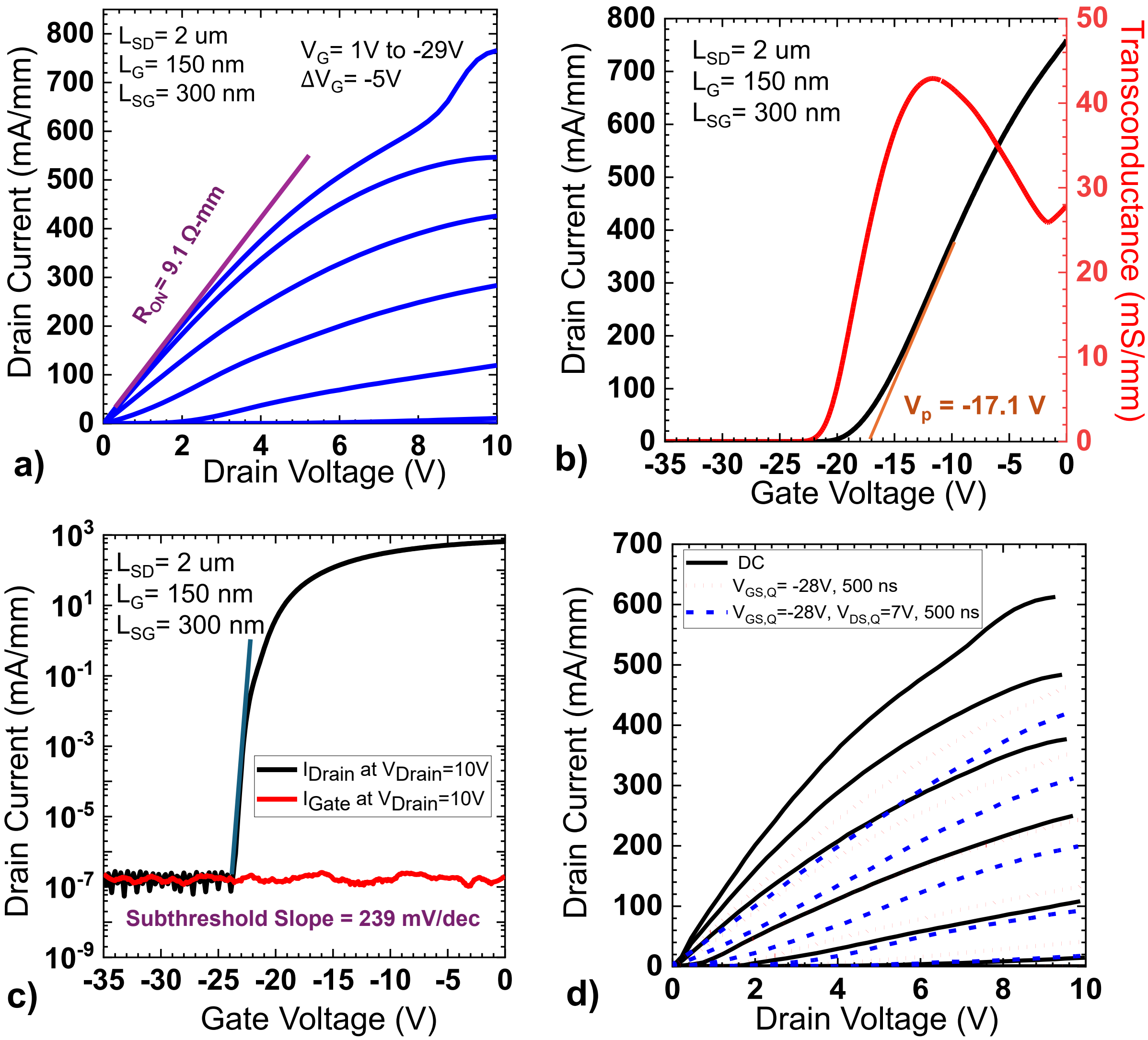

Fig. 4 a) Output IV characteristics b) Transfer IV characteristics c) Log Scale transfer IV characteristics d) Pulsed IV characteristics

The DC output characteristics of the δ-doped β-$Ga_2O_3$ FinFET are shown in Fig. 4(a). The characteristic device has a gate length of 150 nm, source to drain spacing of 2 μm and source to gate spacing of 300 nm. The drain current is normalized to the fin width (33% of the total device periphery with fin width $W_{fin}$ = 100 nm and fin spacing $W_{spacing}$ = 200 nm) rather than the total device perimeter, as conduction in these devices is confined to the fin area, outside which the δ-doped channel is etched away. A maximum drain current of 764 mA/mm was measured at $V_{GS}$ = 1 V and $V_{DS}$ = 10 V, representing the highest reported value to date for β-$Ga_2O_3$ FinFETs. The on-resistance ($R_{ON}$) of the device is 9.1 Ω·mm (27.3 Ω·mm when normalized to device periphery). Despite using a low-power etch process for fin fabrication, the on-resistance ($R_{ON}$) is largely dominated by the resistance of the fin region (see Fig. S1), which is likely attributable to plasma-induced damage during the ICP-RIE etching process[19]. Transfer measurements (Fig. 4(b)) conducted at $V_D$=10V show clear pinch-off with a threshold voltage of ~ -17.1V, subthreshold slope of ~ 239 mV/dec and an $I_{ON}/I_{OFF}$ ratio exceeding $10^9$ with low gate leakage. As mentioned previously, SIMS analysis revealed that the 300 nm buffer layer is unintentionally n-type doped (Si:1x$10^{17}$ $cm^{-3}$) This background doping modifies the electrostatic band profile (Fig. 3 (d-f)), particularly affecting charge depletion behavior and shifts the threshold to larger values. (Fig. 4(b)) shows the transfer characteristics and transconductance with $g_m$ exhibiting a pronounced peak of ~ 43 mS/mm at $V_G$ ~ -11.7 V. Beyond the peak in $g_m$, a substantial roll-off is observed as the gate bias is decreased toward 0 V. This behavior is likely due to reduced mobility in the Si $\delta$-doped channel near the fin sidewalls, resulting from plasma-induced sidewall damage during the fin etch process. At larger reverse gate biases, the modulated channel is located farther from the sidewalls, where the transport properties are less affected by the fin etch-induced damage, resulting in higher $g_m$. Pulsed I–V measurements (Fig. 4(d)) using a 500 ns pulse width reveal a current collapse of approximately 20%, indicative of moderate trapping effects. This collapse is significantly smaller than the severe degradation (~60%) reported in unpassivated recessed-channel β-$Ga_2O_3$ MOSFETs under 200 ns gate turn-on pulsing[25] and is comparable to the gate-lag dispersion (~22%) observed in AlGaO/ β-$Ga_2O_3$ HFETs prior to passivation [26], although it remains higher than the near collapse-free behavior demonstrated in recently optimized β-$Ga_2O_3$ RF FETs [23]. The observed degradation is most plausibly attributed to trap charging at the etched fin sidewalls and at or near the $Al_2O_3$/β-$Ga_2O_3$ gate interface, consistent with fin etching by $BCl_3$-based ICP-RIE and the use of a thin ALD $Al_2O_3$ dielectric. Additional degradation under combined gate and drain quiescent bias further suggests contributions from surface trap states in the gate–drain access regions. Overall, the measured current collapse is encouraging given the >750 mA/mm current density achieved in this fin architecture, while also highlighting the need for improved sidewall and interface passivation for further RF performance optimization.

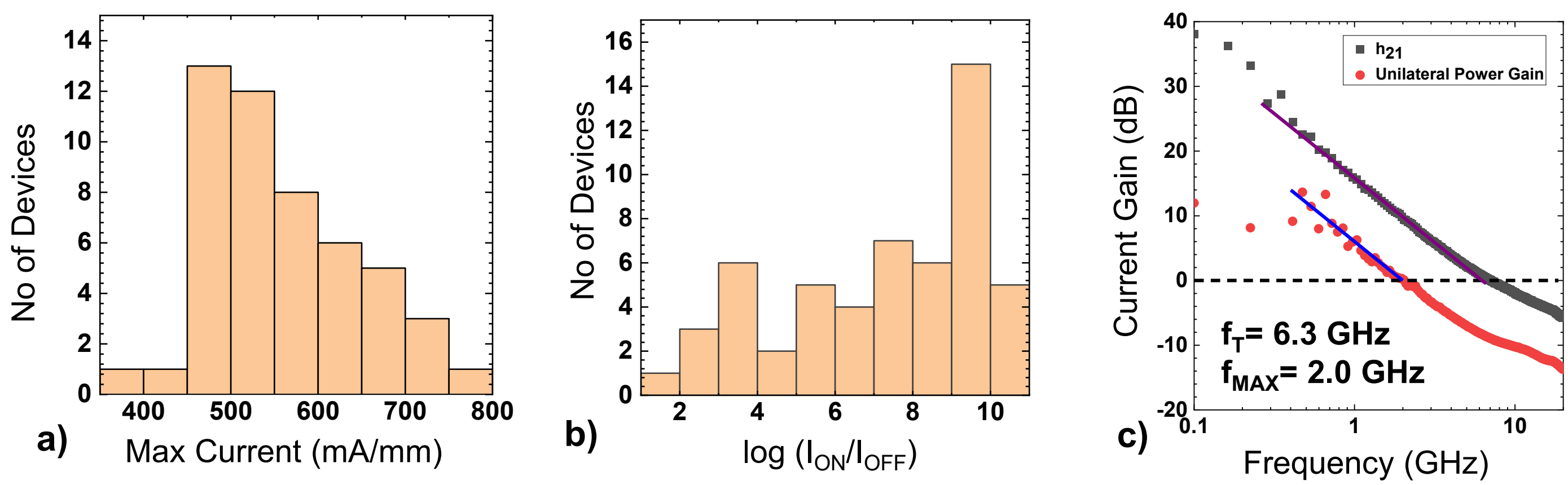

Fig. 5 a)-b) Statistical distribution of device performance showing histograms of *(left)* maximum drain current density and *(center)* log ($I_{ON}/I_{OFF}$) across measured devices, c) Small signal characteristics showing de-embedded high-frequency current gain $|h_{21}|$ and unilateral power gain

The high-frequency small-signal performance (Fig. 5(c)) was measured from 0.1 to 25 GHz using a vector network analyzer. Parasitic pad contributions were removed using open–short de-embedding with dedicated reference structures to extract the intrinsic device response. Figure 5(c) shows the de-embedded short-circuit current gain $|h_{21}|$ and unilateral power gain U. A peak cutoff frequency $f_T$ of 6.3 GHz and a maximum oscillation frequency $f_{MAX}$ of 2.0 GHz were obtained at $V_G = 0$ V and $V_D = 10$ V. The relatively low $f_{MAX}$ is primarily attributed to the high gate resistance associated with the thin (80 nm) Ni gate metallization, which is more resistive than conventional Au-based gates and due to the lack of a T-gate structure. Further improvements in RF performance are therefore expected through reduced gate resistance, improved sidewall passivation to mitigate plasma-induced damage, and optimization of fin geometry to enhance channel mobility and reduce parasitic capacitances. Device yield and process uniformity were evaluated by electrically characterizing a large number of FinFETs across the wafer.

Statistical distribution of the maximum drain current and $I_{ON}/I_{OFF}$ ratio are shown in Fig. 5(a) and Fig. 5(b), respectively. A significant fraction of devices exhibits peak drain currents exceeding 500–700 mA/mm while maintaining $I_{ON}/I_{OFF}$ ratio in the range of $10^8$–$10^{10}$. The simultaneous achievement of high drive current and strong current modulation across multiple devices indicates good uniformity and a high functional yield. Given the challenges associated with submicron β-$Ga_2O_3$ fin fabrication such as plasma-induced damage, sidewall roughness, and interface sensitivity, the observed performance dispersion is attributed primarily to local process variations rather than fundamental limitations of the device architecture. These results suggest that the proposed fabrication process is robust and reproducible, enabling scalable β-$Ga_2O_3$ FinFETs for RF and power applications.

In conclusion, the submicron gated δ-doped β-$Ga_2O_3$ FinFETs were realized through a fabrication flow that integrates $BCl_3$-based fin definition, precise δ-doping placement, PEALD $Al_2O_3$ gate dielectric formation, and Ni-based uniform gate metallization. The resulting transistors exhibit high drive current exceeding 750 mA/mm, $I_{ON}/I_{OFF}$ ratio of $10^9$, with statistical measurements confirming good uniformity and functional yield. RF characterization further shows GHz-scale performance with a $f_T$ of 6.3 GHz, while pulsed I–V measurements indicate moderate trapping behavior and current collapse likely associated with traps at the $Al_2O_3/Ga_2O_3$ MOS fin sidewalls. Collectively, these results demonstrate a scalable β-$Ga_2O_3$ FinFET process capable of delivering competitive DC and RF characteristics, while highlighting opportunities for improved interface and sidewall engineering to advance device performance.

**ACKNOWLEDGEMENTS**

This material is based upon work supported by Army Research Office UWBG RF center under award No. W911NF2520005 and partly by NSF under Grant No. NSF ECCS 2336397. Scanning Electron Microscopy (SEM) was performed at the John M. Cowley center for high resolution microscopy at Arizona State University. We acknowledge the use of facilities within the ASU NanoFab supported in part by NSF program NNCI-ECCS-1542160. C.A.G. acknowledges support from the National Defense Science and Engineering Graduate (NDSEG) Fellowship. This material is based upon work supported by the Air Force Office of Scientific Research under award number FA9550-23F-0014 in the amount of $103,174.25. UB acknowledges the support from ARPA-E Award No. DE-AR0001879, AFOSR (Air Force Office of Scientific Research) under award FA9550-18-1-0479 (Program Manager: Ali Sayir), from NSF under awards ECCS 223102, 2532899 and Coherent II-VI Foundation Block Gift Program.

**Data Availability Statement**

The data that supports the findings of this study are available from the corresponding author upon reasonable request.

# SUPPLEMENTARY INFORMATION

# Sub-Micron Gate β-$Ga_2O_3$ FinFETs with Current Density > 750 mA/mm and $I_{ON}/I_{OFF}$ Ratio > $10^9$

Nabasindhu Das[1], Cameron Gorsak[2], Pushpanshu Tripathi[2], Ankita Kashyap[1], Noor Jahan Nipu[3], Advait Gilankar[1], Hari Nair[2], Uttam Singisetti[3], Trevor Thornton[1], Nidhin Kurian Kalarickal[1]

[1] *School of Electrical, Computer and Energy Engineering, Arizona State University, Arizona, USA*

[2] *Department of Materials Science & Engineering, Cornell University, Ithaca, New York, USA*

*3Department of Electrical Engineering, University at Buffalo, New York, USA*

*ndas11@asu.edu, Phone: (602)-813-6289*

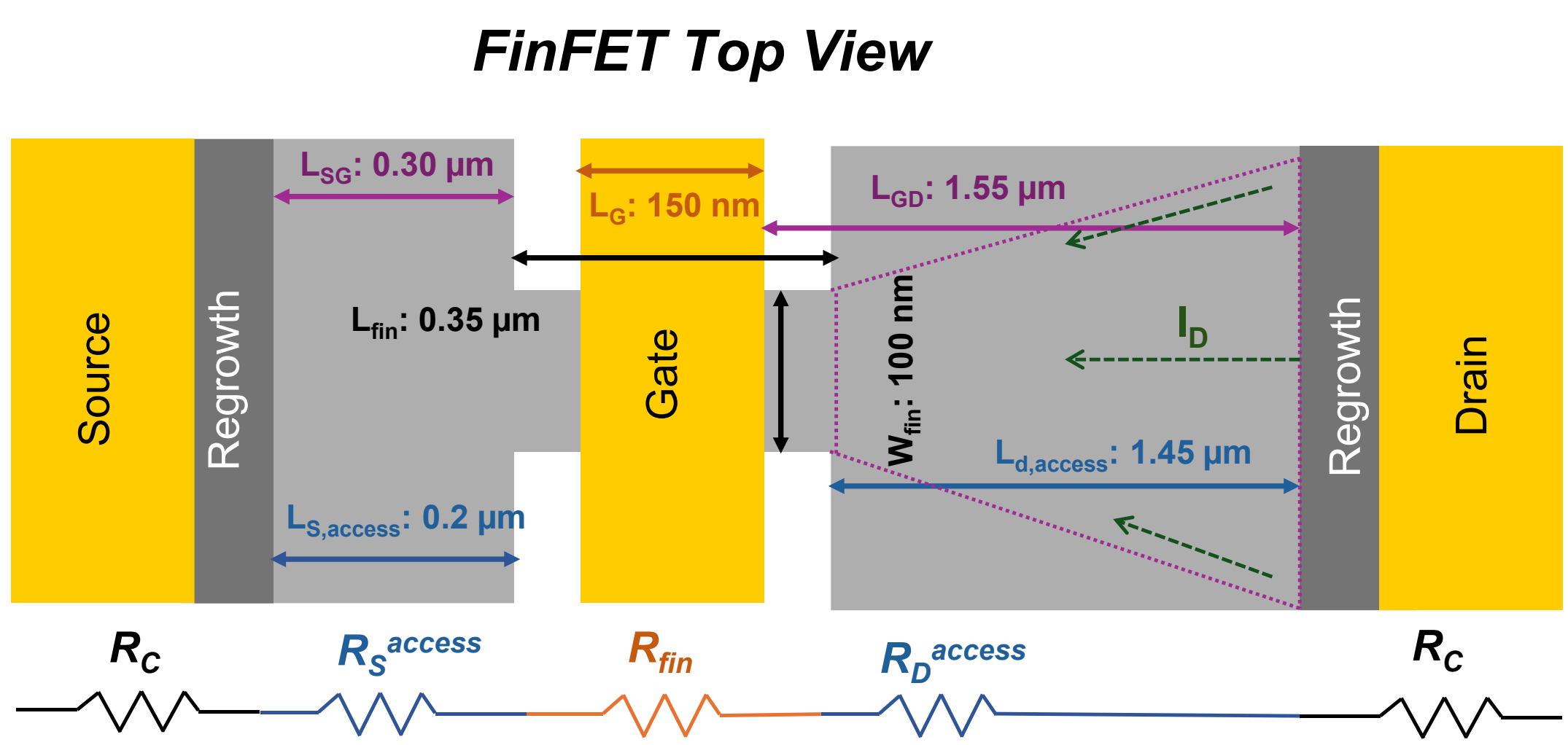

Fig. S1 Top-view schematic of the β-$Ga_2O_3$ FinFET showing the gate, source/drain regrowth regions, fin and access lengths, and corresponding resistance components.

The following summarizes the calculations used to separate the total on-resistance of the FinFET device into contact resistance, source/drain access resistance, and fin resistance. The values are reported in ohm millimeter. The total on-resistance normalized to device periphery is 27.3 Ω.mm.

## Calculation of Resistance

The sheet and contact resistance for the device to the delta-doped epitaxial layer are extracted from fabricated TLMs with different spacings where the sheet resistance is extracted from slope and contact resistance from intercept.

The total on resistance of the FinFET is given by:

$$R_{on} = 2R_c + R_s^{access} + R_d^{access} + R_{fin}$$

where $R_c$ is the contact resistance, $R_s^{access}$ and $R_d^{access}$ are the source and drain access resistance and $R_{fin}$ is the fin resistance.

1. **Contact Resistance**
   Considering both source and drain contacts are equivalent, the net contact resistance is $R_{c,net} = 2 \times R_c$ where $R_c$ is the TLM extracted contact resistance. The contact resistance value is 0.16 Ω.mm.

2. **Access Region Resistance**
   In the FinFET geometry, the source and drain access regions cannot be treated as rectangular resistors of constant width because the current flowing through the wider regrown region must converge into the narrower fin. Treating the current path as a linear trapezoid extending from the fin pitch ($p$) to the fin width (W) for access region length ($L_A$) for a fin fill factor $D = \frac{W_{fin}}{p}$ the width normalized access resistance is $R_{acess,norm} = R_{sh} \times L_A \times F_{crowd}$ where $F_{crowd} = \frac{ln(1/D)}{1-D}$ is the current crowding correction to existing rectangular expression $R_{sh} \times L_A$. This expression is applied separately to the source and drain using their respective access lengths. The source and drain access resistances $R_s^{access}$ and $R_d^{access}$ are estimated to be 0.69 Ω.mm and 5.03 Ω.mm respectively.

3. **Fin Resistance**
   For a fin array, only a fraction of width is occupied by fins that conduct. The fin fill factor is the fin width divided by the pitch: $D = \frac{W_{fin}}{p}$. For a 100 nm fin with 200 nm gap between neighboring fins, the pitch is 300 nm and the fill factor is $1/3^{rd}$. The width normalized fin resistance is therefore: $R_{fin,norm} = \frac{R_{sh}L_{fin}}{D}$. $R_{fin,norm}$ represents the theoretical resistance estimated from the sheet resistance and device geometry, whereas $R_{fin,experimental}$ is extracted from the measured on-resistance after subtracting the contact and access resistance contributions. $R_{fin,norm}$ is 2.21 Ω.mm and $R_{fin,experimental}$ is 21.42 Ω.mm. The substantially higher experimental value highlights the impact of plasma-induced damage from the fin etching process.

The table below summarizes the percentage contribution and extracted values of the contact resistance ($R_{c,net}$), source/drain access resistances ($R_s^{access}$, $R_d^{access}$) and calculated and experimental intrinsic fin resistances ($R_{fin}$) as explained above.

| Region | Percentage | Resistance |
|---|---|---|
| $R_{c,net}$ | 0.5% | 0.16 Ω.mm |
| $R_s^{access}$ | 2.5% | 0.69 Ω.mm |
| $R_d^{access}$ | 18.4% | 5.03 Ω.mm |
| $R_{fin,norm}$ | 8% | 2.21 Ω.mm |
| $R_{fin,experimental}$ | 78.4% | 21.42 Ω.mm |